# Bibliometric evaluation of research performance: where do we stand?[1]

*Giovanni Abramo*

Laboratory for Studies in Research Evaluation
Institute for System Analysis and Computer Science (IASI-CNR)
National Research Council of Italy
Via dei Taurini, 19 - 00185 Roma – Italy

**Abstract**

This work provides a critical examination of the most popular bibliometric indicators and methodologies to assess the research performance of individuals and institutions. The aim is to raise the fog and make practitioners more aware of the inherent risks in do-it-myself practices, or cozy out-of-the-shelf solutions to the difficult question of how to evaluate research. The manuscript also proposes what we believe is the correct approach to bibliometric evaluation of research performance.

---

## 1. Introduction

In the current knowledge-based economy of a globalized world, research-based innovations are increasingly becoming sources of competitive advantage both at industry and nation levels. Therefore, improving the effectiveness and efficiency of the domestic scientific and technology infrastructure ranks among the top priorities in the policy agenda of many governments. Universities and public research institutions, being the heart of this infrastructure, play a vital role in the generation and transmission of new knowledge and discoveries, and consequently an increasingly decisive role in industrial competitiveness, economic growth and employment. At the same time, the rising costs of research and the tight restrictions in public budgets, call for the adoption of more efficient systems of resource allocation. To stimulate continuous improvement, enhance accountability and better manage public funds, a rising number of nations have implemented research assessment exercises. Alongside, many of them have shifted from conventional funds allocation remunerating institutional size and type of research, to one based on research performance. The assessment exercises serve towards five principal objectives, adopted in whole or in part by the governments concerned: i) stimulation of greater production efficiency; ii) selective funding allocations; iii) reduction of information asymmetry between supply and demand in the market for knowledge; iv) informing research policy and strategy; and last but not least v) demonstration that investment in research is effective and delivers public benefits. As a consequence, a demand for increasingly precise indicators of research performance and methods to assess it has blown. Over recent years, scientometricians have proposed different methods of evaluation and a myriad of indicators and their variants, and the variants of the variants (scientometricians are now running out of alphabet and subscript characters to name all the new indicators/variants). The proliferation of proposals has actually generated a type of disorientation among decision makers, no longer able to discriminate the pros and cons of the various indicators and methods for planning an actual evaluation exercise. The proof of this is the increasing number of expert commissions and working groups at institutional, national and supranational levels, formed to deliberate and recommend on this indicator, that set of indicators, and this or that methodology to assess performance. Performance ranking lists at national and international levels are published with media fanfare, influencing opinion and practical choices. Unfortunately, the impression of the current author is that these rankings of scientific performance, produced by "non-bibliometricians" (THE 2016; SJTU 2016; QS 2016; etc.) and even by bibliometricians (University of Leiden, SCImago, etc.), are largely based on what can easily be counted rather than "what really counts".

In this work, I provide a critical examination of the most popular bibliometric indicators and methodologies to assess the research performance of individuals and institutions. The aim is to raise the fog and make practitioners more aware of the inherent risks in do-it-myself practices, or cozy out-of-the-shelf solutions to the difficult question of how to evaluate research. This paper does not say anything new or different from what can be found here and there in my previous works. I apologize then if the reference list in the end contains so many self-citations. I hope the reader finds them worth reading, anyway. What is new about this work, is the systemic overview of where we stand in terms of bibliometric evaluation of research. I will be critical and straightforward in commenting current practices of research evaluation, as could be expected from somebody whose "ideas differ fundamentally from mainstream scientometric thinking" (Waltman, 2016). I have also to recognize that the ideas that I am going to present are the outcome of a several years' joint research effort at the research laboratory that I co-founded with colleague Ciriaco Andrea D'Angelo. Needless to say, I am the only one responsible for these ideas, although most of the credit for the underlying work goes to all the research staff and Phd students that have been or are still member of the lab. Of course, I will not limit myself to criticize current practices, which would be a hollow exercise, but I will also propose what I believe is the correct approach to bibliometric evaluation of research performance. The next section of the paper deals with bad practices and invalid indicators of research performance.

The third section proposes what we believe at our lab is the correct approach to research evaluation. The fourth section draws the conclusions.

### 2. Invalid bibliometric indicators and rankings

Until now, the bibliometrics literature has proposed indicators and methods for measuring research performance that are largely inappropriate from a microeconomics perspective. In the following, I will critically analyze the most popular ones. Perhaps the most striking example is the indicator of research productivity. Bibliometricians have become accustomed to define productivity as the number of publications per researcher, distinguishing it from impact, which they measure by citations. Honestly, I am not able to date back the scholar who first introduced the above definition, but already in 1926 Alfred J. Lotka used the number of publications in his milestone work (Lotka 1926) where he presented what it is now known as the Lotka's law or research productivity. Unfortunately, from an economic standpoint, such definition makes little sense. It would be acceptable only if all publications had the same value or impact, but that could not be further from the truth. It is like saying that two automobile manufacturers, producing respectively Fiat 500 cars and Ferrari 488 cars, have the same productivity because they produce the same number of automobiles per day, all production factors being equal; or, it is like measuring the GDP of a country by counting the number of widgets produced, regardless of their market value.

Another category of invalid indicators is the one represented by citation size-independent indicators based on the ratio to publications, whose most popular representative is the mean normalized citation score or MNCS. The MNCS is claimed as an indicator of research performance, measuring the average number of citations of the publications of an individual or institution, normalized for subject category and publication year (Waltman et al., 2011). Similarly, the share of individual or institutional publications belonging to the top 1% (10%, etc.) of 'highly cited articles' (HCAs), compared with other publications in the same field and year, is considered another indicator of research performance. Abramo and D'Angelo (2016a; 2016b) object to it. Given two universities of the same size, resources and research fields, which one performs better: the one with 100 articles each earning 10 citations, or the one with 200 articles, of which 100 with 10 citations and the other 100 with five citations? A university with 10 HCAs out of 100 publications, or the one with 15 HCAs out of 200 publications? In both examples, by MNCS or proportion of HCAs, the second university performs worse than the first one (25% lower). But clearly, using common sense, the second is in both cases the better performer, as it shows higher returns on research investment (50% better). Basic economic reasoning confirms that the better performer under parity of resources is the actor who produces more; or under parity of output, the better is the one who uses fewer resources. Indeed the MNCS, the proportion of HCAs, and all other size-independent indicators based on the ratio to publications are invalid indicators of performance, because they violate an axiom of production theory: as output increases under equal inputs, performance cannot be considered to diminish. Indeed an organization (or individual) will find itself in the paradoxical situation of a worsened MNCS ranking if it produces an additional article, whose normalized impact is even slightly below the previous MNCS value.

Another world renown performance indicator is the *h-index*, proposed in 2005 by the Argentine American physicist, J.E. Hirsch (Hirsch, 2005). The h-index represents the maximum number *h* of works by a scientist that have at least *h* citations each. Hirsch's intuitive breakthrough was to represent with a single whole number a synthesis of both the quantity and impact of a scientist's portfolio of work. However, the h-index and most of its variants ignore the impact of works with a number of citations below *h* and all citations above *h* of the h-core works. Furthermore, the h-index fails to field-normalize citations, and to account for the number of co-authors and their order in the byline. Last but not least, because of the different intensity of publications across fields, productivity rankings need to be carried out by field (Abramo & D'Angelo, 2007), when in reality there is a human tendency to compare h-indexes for researchers across different fields. Each one of the proposed h-variant indicators tackles one of the many drawbacks of the h-index while leaving the others unsolved, so none can be considered completely satisfactory.

A trend we are all witnessing is the annual publication of international rankings of individual research institutions. Before forging their perceptions or making any decisions based on them, decision makers should pay a special attention to the "supposed performance indicators" underlying such rankings. For example, the CWTS Leiden Rankings (2016) are based on such invalid indicators as the total number of publications; the proportion of HCAs; and, till 2015, the MNCS. Similar drawbacks are embedded in the SCImago Institutions Ranking (2016) by their main indicator, the Normalized Impact, measuring the ratio between the average scientific impact of an institution and the world average impact of publications of the same time frame, document type and subject area. I do not further consider any of the many annual world institutional rankings produced by non-bibliometricians (THE 2016; SJTU 2016; QS 2016; etc.). In these rankings, the performance indicators are given different weight in determining the position of universities. However, their use presents distortions both due to the lack of field-standardization and to strong size-dependency. The SJTU-Shanghai Academic Ranking of World Universities, for example, is notorious for the fact that over 90% of the performance result depends on university size. No surprise then if these non-scientific rankings are given more coverage in popular and promotional media, while are heavily criticized in the scientific press.

As for national comparative research performance exercises of universities and institutions, according to Hicks (2012), there are at least 15 nations (China, Australia, New Zealand, 12 EU countries) that conduct them regularly and link the results to public financing. The recent development of bibliometric techniques has led various governments to introduce bibliometrics, where applicable, in support for more traditional peer review. In the United Kingdom the 2014 Research Excellence Framework (REF), which replaced the peer-review RAE (Research Assessment Exercise) series, was the first UK informed peer-review exercise where the assessment outcomes were a product of expert review informed by citation information and other quantitative indicators. The problem with peer-review or informed peer-review national-scale evaluation exercises is that they must necessarily be based on a subpopulation of products, for reasons of time and costs. Differently, if the evaluation exercise is based on bibliometric techniques and indicators this limitation no longer occurs. The bibliometric approach offers at least two clear advantages: i) it avoids the distortion of performance due to inefficient selection of products for evaluation, on the part of individual scientists and of their institutions; and ii) it avoids distortions due to evaluating only a part of the research product. Abramo et al. (2009) first quantified these distortions for the case of Italy's first research assessment exercise VTR 2004-2006. Abramo et al. (2014) in particular, have estimated the error in the selection of products for the hard sciences: the results indicate a worsening the maximum score achievable by 23% to 32%, compared to the score from an efficient selection. Abramo et al. (2010b) conducted also a sensitivity analysis of performance rankings to the share of research product evaluated. In terms of accuracy, robustness, validity, functionality, time and costs, the superiority of bibliometrics compared to peer review has been demonstrated by Abramo and D'Angelo (2011). Still, there is a strong resistance by governments and part of the academic community, to substitute peer review with bibliometrics, where applicable, in large-scale evaluations.

3. **The correct approach to bibliometric evaluation of individuals and organizations**

Together with my colleague, Ciriaco Andrea D'Angelo, we have formulated a proxy of the quintessential indicator of efficiency of any production unit, productivity. We have been applying it for several years to measure and rank the performance of Italian academics and research institutions. We devoted a specific work to provide an operative definition of our proxy indicator of productivity and the method to apply it (Abramo and D'Angelo, 2014). In this section, I will report the main characteristics of it, while I refer the reader to the above mentioned original paper for further details.

Research organizations are no different from any other production systems. They use resources (production factors) to produce output (new knowledge). The microeconomic theory of production describes the relation between the two by the well known production function: Q = F(K, L), where *Q* is the output, *L* is labor and *K* are all production factors other than *L*. Because of the nature of research systems, to measure productivity one needs adopt a few simplifications and assumptions both on the output and the input side. As for the first, new knowledge, i.e. research output, is intangible. Because one can measure only what is quantifiable, as a proxy of output bibliometricians use publications (indexed in such bibliometric databases as WoS or Scopus). An immediate consequence of this is that in those disciplines (mainly arts and humanities) where the coverage or research output by bibliometric databases is limited, bibliometric techinques cannot be applied to research evaluation. Publications have different value or impact on scientific advancement, which bibliometricians approximate with citations. It must be noted that the journal impact factor should never be used as a substitute of or in combination with citations, unless the citation window is extremely short (Abramo et al. 2011; Abramo, D'Angelo & Di Costa, 2010a, Levitt & Thelwall, 2011; Stern, 2014; Abramo & D'Angelo, 2016c). Because citation behavior varies by field, we standardize the citations for each publication with respect to the average of the distribution of citations for all the cited publications indexed in the same year and field.[2] The intensity of publication also varies by field, a prerequisite then of any distortion-free performance assessment is to classify each researcher into a single field (Abramo et al. 2013b). Furthermore, research projects frequently involve a team of researchers, which is registered in the co-authorship of publications. In this case, we account for the fractional contributions of scientists to outputs, which is sometimes further signaled by the position of the authors in the list of authors.

On the side of production factors, there are again difficulties in measure that lead to inevitable approximations. The identification of production factors other than labor and the calculation of their value and share by fields is formidable (consider quantifying value of accumulated knowledge or scientific instruments shared among units). In many countries, even the identification of the researchers in each institution may reveal a formidable task, not to talk about their classification into research fields. In Italy, we gain advantage from a database maintained by the Ministry of Education, University and Research, which indexes all academics by their affiliation, academic rank, and field of research. The latter characteristic seems unique to the Italian higher education system, in which each professor is classified as belonging to a single research field. These formally-defined fields are called "Scientific Disciplinary Sectors" (SDSs): there are 370 SDSs, grouped into 14 "University Disciplinary Areas" (UDAs).

Because of lack of information on the capital *K* available to each individual or unit, the measure of total factor productivity is generally impossible. Thus, an often-necessary assumption is that the resources available to individual/units within the same field are the same. A further assumption, again unless specific data are available, is that the hours devoted to research are more or less the same for each individual. Finally, as occurs for output, the value of researchers is not undifferentiated and this is reflected in the different cost of labor, which varies among research staff, both within and between units. If cost of labor is available, one should normalize output by it.

When measuring research productivity, the specifications for the exercise must also include the publication period and the "citation window" to be observed. The choice of the publication period has to address often contrasting needs: ensuring the reliability of the results issuing from the evaluation, but also permitting conduct of frequent assessments. For the most appropriate publication period to be observed see Abramo et al. (2012a), while for the citation window that optimizes the tradeoff between accuracy of rankings and timeliness of the evaluation exercise, see Abramo et al. (2012b).

We have named our indicator representing the proxy of the average yearly productivity over a period of time, Fractional Scientific Strength, or FSS. At the individual researcher level *R,* we then measure $FSS_R$, accounting for the cost of labor, in the following way:

---

[2] Abramo et al. (2012c) demonstrated that the average of the distribution of citations received for all cited publications of the same year and field is the most effective scaling factor.

$$FSS_R = \frac{1}{w_R} \cdot \frac{1}{t} \sum_{i=1}^{N} \frac{c_i}{\bar{c}} f_i$$

[1]

Where:
$w_R$ = average yearly salary of the researcher
$t$ = number of years of work by researcher in period under observation
$N$ = number of publications by researcher in period under observation
$c_i$ = citations received by publication, $i$
$\bar{c}$ = average of distribution of citations received for all cited publications in same year and subject category of publication, $i$
$f_i$ = fractional contribution of researcher to publication, $i$.

The fractional contribution equals the inverse of the number of authors in those fields where the practice is to place the authors in simple alphabetical order but assumes different weights in other cases. For the life sciences, widespread practice in Italy is for the authors to indicate the various contributions to the published research by the order of the names in the listing of the authors. For the life science SDSs, we give different weights to each co-author according to their position in the list of authors and the character of the co-authorship (intra-mural or extra-mural) (Abramo et al. 2013c). If the first and last authors belong to the same university, 40% of the citation is attributed to each of them, the remaining 20% is divided among all other authors. If the first two and last two authors belong to different universities, 30% of the citation is attributed to the first and last authors, 15% of the citation is attributed to the second and last authors but one, the remaining 10% is divided among all others.[3]

Operationally, in the Italian case, beginning from the raw data of the WoS, and applying a complex algorithm to reconcile the author's affiliation and disambiguation of the true identity of the authors, each publication is attributed to the author(s) that produced it (D'Angelo et al. 2011). Thanks to this algorithm we can produce rankings of research productivity at the individual level, on a national scale. Based on the score of $FSS_R$ we obtain, for each SDS, a ranking list expressed on a percentile scale of 0–100 (worst to best), or as the ratio to the average productivity of all Italian colleagues of the same SDS with productivity above zero.[4] This allows to compare performance of all Italian academics regardless of the SDS they belong to.

In multi-field organizational units (i.e. disciplines, departments, universities, regions, nations), where there are researchers that belong to different fields, we are presented with the problem of how to aggregate productivity measures for researchers from the various fields. We have seen that the performance of the individual researchers can be expressed in percentile rank or standardized to the field average. We avoid averaging percentile ranks of the researchers. Thompson (1993) warns that percentile ranks should not be added or averaged, because percentile is a numeral that does not represent equal-interval measurement. Further, percentile rank is also sensitive to the size of the fields and to the performance distribution. We resort then to standardized $FSS$, which accounts for the extent of difference between productivities of the individuals. In formula, the productivity $FSS_U$ over a certain period for a multi-field research unit $U$:

$$FSS_U = \frac{1}{RS} \sum_{j=1}^{RS} \frac{FSS_{R_j}}{\overline{FSS_R}}$$

[3]

---

[3] Different practices may occur in other countries, whereby the fractional contributions may be adapted accordingly.
[4] Abramo et al. (2013) demonstrated that the average of the productivity distribution of researchers with productivity above 0 is the most effective scaling factor to compare the performance of researchers of different fields.

Where:

$RS$ = research staff of the unit, in the observed period;

$FSS_{R_j}$ = productivity of researcher $j$ in the unit;

$\overline{FSS_R}$ = national average productivity of all productive researchers in the same SDS of researcher $j$.

## 4. Conclusions

The great majority of the bibliometric indicators and the rankings based on their use present two fundamental limits: lack of normalization of the output value to the input value, and absence of classification of scientists by field of research. Without normalization there cannot be any measure of productivity, which is the quintessential indicator of performance in any production unit; without providing field classification of scientists, the rankings of multi-field research units will inevitably be distorted, due to the different intensity of publication across fields. An immediate corollary is that it is impossible to correctly compare productivity of institutions at international levels. In fact, there is no international standard for classification of scientists and, we are further unaware of other nations that classify their scientists by field at domestic level, apart from Italy and the Scandinavian countries. This obstacle can in part be overcome by indirectly classifying researchers according to the classification of their scientific production into WoS or Scopus categories, and then identifying the predominant category. Fractional Scientific Strength (FSS) is a proxy indicator of productivity permitting measurement at different organizational levels. Both the indicator and the related methods can certainly be improved, however they do make sense according to economic theory of production. Other indicators and related rankings, such as the simple number (or fractional counting) of publications per research unit, or the average normalized impact, cannot alone provide evaluation of performance - however they could assume meaning if associated with a true measure of productivity. In fact, if a research unit achieves average levels of productivity this could result from average production and average impact, but also from high production and low impact, or the inverse. In this case, knowing the performance in terms of number of publications and average normalized impact would provide useful information on which aspect (quantity or impact) of scientific production to strengthen for betterment of production efficiency.

While it may be debatable whether it was Albert Einstein or William Cameron that coined the saying, '*Not everything that can be counted counts, and not everything that counts can be counted*', no one doubts its pertinence and extraordinary importance in the field of scientometrics. Anyone involved in research evaluation should always keep in mind that pill of wisdom, and count only what counts.